\documentclass{article}
\usepackage{spconf,amsmath,graphicx,hyperref}
\usepackage[capitalize]{cleveref}
\usepackage[ruled,vlined,linesnumbered,lined,algoruled]{algorithm2e}
\usepackage{xcolor}
\usepackage{booktabs}

\title{Source Separation for A Cappella Music}
\name{Luca A. Lanzendörfer$^*$ \qquad Constantin Pinkl$^*$ \qquad Florian Grötschla
\thanks{$^*$Equal contribution}}
\address{ETH Zurich}

\begin{document}

\newcommand{\model}{SepACap}

\maketitle

\begin{abstract}
In this work, we study the task of multi-singer separation in a cappella music, where the number of active singers varies across mixtures. To address this, we use a power set-based data augmentation strategy that expands limited multi-singer datasets into exponentially more training samples. To separate singers, we introduce SepACap, an adaptation of SepReformer, a state-of-the-art speaker separation model architecture. We adapt the model with periodic activations and a composite loss function that remains effective when stems are silent, enabling robust detection and separation. Experiments on the JaCappella dataset demonstrate that our approach achieves state-of-the-art performance in both full-ensemble and subset singer separation scenarios, outperforming spectrogram-based baselines while generalizing to realistic mixtures with varying numbers of singers.
\end{abstract}


\section{Introduction}
The field of music source separation has recently seen rapid progress~\cite{defossez2019demucs,rouard2023hybrid,luo2023music,wang2023mel}, primarily focusing on separating different instruments from each other by learning spectral masks that isolate the desired source from the mixture. In contrast, separating multiple singers in purely vocal (a cappella) recordings remains underexplored. A cappella ensembles range from small groups (duets and quartets) to chamber choirs, often organized by part (soprano, alto, tenor, and bass), as well as subset variants~\cite{petermann2020deep,chen2022improving,jeon2023medleyvox}. These recordings exhibit dense harmonic overlap, frequent unison/octave doubling within a part, voice crossing, vibrato and portamento, tightly aligned consonant onsets, and breath and sibilant noise. These properties reduce timbral diversity and make source separation more challenging than instrument-wise separation~\cite{cuesta2019framework}. Contemporary a cappella may also include vocal percussion or beatboxing, further increasing spectral overlap despite all sources being human voices~\cite{delgado2019new,yong2023phoneme}.

Despite applications in transcription, remixing, and choir analysis, singer separation has received relatively little attention. A major bottleneck is the limited availability of multi-singer datasets. These constraints make it difficult to train large neural network-based separation models and data augmentation strategies are needed to obtain more training samples.

In this work, we present \model{}, an adaptation of the recently proposed SepReformer~\cite{shin_separate_2025}, a state-of-the-art speaker separation model, for the a cappella setting where the number of active singers may vary across mixtures. We adapt the model by (i) introducing periodic activations~\cite{ziyin2020neuralsnake} as we find them to perform better compared to the default ReLU activations; (ii) replacing the default training loss with a silence-aware composite objective that remains well-defined when stems are absent, combining waveform, multi-scale mel, and multi-resolution spectral losses; and (iii) coupling the model with a power set-based data augmentation scheme that creates mixtures for all subsets of stems, enabling joint separation and detection of active singers. We evaluate \model{} on JaCappella~\cite{nakamura2023jacappella} in two scenarios: all-stems and subset-stems, and report both separation quality and silent-stem suppression quality, along with detection metrics. Across these settings, \model{} outperforms strong spectrogram-masking and waveform baselines, achieving state-of-the-art results in full-ensemble separation while markedly reducing bleed-through and correctly outputting silence for inactive stems.

Our contributions can be summarized as follows:
\begin{itemize}
    \item We propose \model{}, an a cappella source separation model achieving state-of-the-art performance on multi-singer separation while operating in the waveform domain.
    \item To increase available training data, and to enable model generalizability to subset singer separation, we use a power set-based data augmentation method that transforms a standard multi-singer dataset into exponentially more training samples, including cases with absent singers, enabling more robust separation performance.
    \item We extend the loss function to allow for stable training with empty stems, ensuring that models operating on the waveform learn to handle silent signals in the data.
\end{itemize}

\section{Related Work}

\noindent \textbf{Audio separation.} 
Research in music source separation has advanced along two principal directions. Spectrogram-based methods typically predict time-frequency masks applied to the mixture. 
Early CNN- and RNN-based systems have been surpassed by Transformer architectures that better capture spectral structure. 
For example, Band-Split RoFormer (BS-RoFormer)~\cite{su2021roformer} introduced subband projections and interleaved Transformers, achieving first place in the Sound Demixing Challenge 2023~\cite{fabbro2023sound}, while its successor, Mel-Band RoFormer~\cite{wang2023mel}, replaced empirical band splits with mel-scale projections for improved perceptual resolution. 
Similarly, TF-Locoformer~\cite{tfLocoformer} combined local convolutional and global Transformer modeling in the time-frequency domain. 
In contrast, waveform-based approaches estimate sources directly without masks. 
Conv-TasNet~\cite{luo2019conv} pioneered learnable time-domain encoders and decoders, Demucs~\cite{defossez2019demucs} extended this with a U-Net-like architecture, and SepReformer~\cite{shin2024separate} combined global and local Transformer blocks with an \emph{early split} into speaker-specific features, enabling efficient long-sequence processing. 
Together, these methods highlight the effectiveness of both spectrogram- and waveform-based modeling for music separation.

\smallskip
\noindent \textbf{Separation of singers.} Compared to instrument separation, singer separation has received less attention despite its relevance for applications such as transcription, remixing, and choir analysis. Ensemble singing is especially challenging due to overlapping harmonics between singers. Therefore, most work in this domain focuses on filtering out the lead vocal. 
The authors of JaCappella~\cite{nakamura2023jacappella} use a variety of models that are capable of separating singers which are singing at various pitch levels. They compare the performance of CrossNet-Open-Unmix (X-UMX)~\cite{sawata2021all}, DPT-Net~\cite{chen2020dualpathtransformernetworkdirect}, and MRDLA~\cite{nakamura2021time} on JaCappella. It should be noted that all tested approaches rely on all singers being present in the music mixture.

\smallskip
\noindent \textbf{Datasets for singer separation.} Compared to other separation domains, singer separation datasets are highly limited. JaCappella~\cite{nakamura2023jacappella} provides 6 vocal stems and finger snapping, totaling 34 minutes of audio across 35 songs in diverse genres such as jazz, punk rock, bossa nova, popular, reggae and enka. In addition to JaCappella, other a cappella datasets exist but are less diverse, less clean, or smaller in size: Dagstuhl ChoirSet~\cite{rosenzweig2020dagstuhl} with 55 minutes of music from 2 classical songs performed by a choir of 16 singers. It includes the stems soprano, alto, tenor, and bass. Another dataset is ESMUC Choir~\cite{cuesta2022data}, comprising three songs, one each from the pop, folk, and classical choral genres, totalling approximately 31 minutes of audio. Both datasets suffer from bleed-through between singers and limited coverage of stems.
\section{Methodology}
\label{sec:method}

\textbf{Model.} To separate arbitrary singers in an a cappella setting, we introduce \model{}, an adaptation of SepReformer~\cite{shin_separate_2025}, a recently proposed state-of-the-art speaker separation model. We found that updating the existing ReLU activation functions to the SNAKE activation~\cite{ziyin2020neuralsnake} markedly increased model performance. Additionally, we changed the training objective from SNR-based training to a composite time and time-frequency loss.
Many separation models from the speech domain use a variant of signal-to-noise ratio (SNR) or signal-to-distortion ratio (SDR) as their loss during training~\cite{shin_separate_2025, subakan2021attentionneedspeechseparation, zhao2023mossformer}, such as the scale invariant SDR (SI-SDR)~\cite{le2019sdr} defined as
\begin{align}
    \text{SI-SDR}&:= 10\log_{10}\frac{\|e_{target}\|^2}{\|e_{\text{res}}\|^2}= 10 \log_{10}\frac{\left\|\frac{\hat s^Ts}{\|s\|}s\right\|^2}{\left\|\frac{\hat s^Ts}{\|s\|}s-\hat s\right\|^2},
\end{align}
where $s$ is the target signal and ${\hat s}$ is the predicted signal.

The problem when using SI-SDR as a loss function is that, for stems without a signal, SI-SDR provides no informative gradient, making it ineffective as a training objective in such cases.
Therefore, we utilize a different class of losses that provide a similar information level as the SI-SDR loss, but do not rely on a signal always being present. We propose to use a combination of three different losses. We use an L1 loss on the waveform as well as a multi-scale Mel loss, which measures the L1 distance between log-mel spectrograms. We additionally use a spectral loss, which combines L1 losses on magnitude and log-magnitude STFT features to capture spectral consistency across resolutions. This loss combination has been effectively applied in the audio compression domain~\cite{kumar2023high}. We find that this loss combination works well and demonstrate its successful transfer to the task of a cappella separation. 

\smallskip
\noindent \textbf{Data Augmentation.}
\begin{figure}
    \centering
    \includegraphics[width=1.0\linewidth]{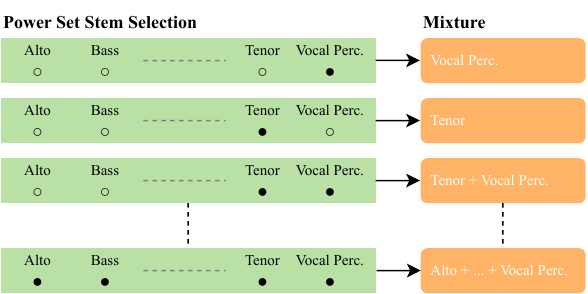}
    \caption{Power-set based data augmentation. For each clip, we construct mixtures by selecting subsets of individual stems (e.g., soprano, alto, tenor) and summing the corresponding stems. This procedure generates all $2^n$ possible combinations of the $n$ available stems, yielding a diverse set of mixtures for training and evaluation.}

    \label{fig:dataug}
\end{figure}
Since our model is designed to handle arbitrary subsets of sources, we formalize the training data generation as follows. 
Let $\mathcal{S} = \{s_1, s_2, \dots, s_n\}$ denote the set of $n$ available stems for a given clip. 
For each non-empty subset $A \subseteq \mathcal{S}$, we construct a mixture 
${
x_A(t) \;=\; \sum_{s_i \in A} s_i(t),
}$
where $s_i(t)$ denotes the time-domain waveform of stem $i$. 
This procedure enumerates the power set $\mathcal{P}(\mathcal{S})$, producing $2^n-1$ distinct mixtures per clip (excluding the empty set). 
 Further, we segment the audio into short fixed-length snippets, which both increases the number of training samples and reduces memory requirements during training.

\section{Experiments}

\begin{table}[t!]
  \centering
  \setlength{\tabcolsep}{6pt}
  \renewcommand{\arraystretch}{0.95}

  \begin{tabular}{lccc}
    \toprule
    Stem &  L1 &  Mel & STFT \\
    \midrule
    Alto        & 9.0 & 6.4 & -15.2 \\
    Bass        & 19.6 & 16.5 & -34.1 \\
    Lead Vocal  & 7.0 & 5.3 & -36.1 \\
    Soprano     & 6.1 & 2.9 & -27.1 \\
    Tenor       & 9.6 & 6.2 & -31.5 \\
    Vocal Perc. & 14.5 & 3.8 & -30.1 \\
    \bottomrule
  \end{tabular}
  \caption{Loss ablation study of \model{} training solely with L1, Mel, or STFT loss, respectively. We report per-stem SDRi on the JaCappella test split where all stems are present.}
  \label{tab:sisdr_ablations}
\end{table}

\begin{table}[t!]
  \centering
  \setlength{\tabcolsep}{4pt}
  \renewcommand{\arraystretch}{0.95}

  \begin{tabular}{lccc|ccc}
    \toprule
    & \multicolumn{3}{c}{SI-SDRi (dB)$\uparrow$} & \multicolumn{3}{c}{RMS (dBFS)$\downarrow$} \\
    \cmidrule(lr){2-4}\cmidrule(lr){5-7}
    Stem &  L1 & Mel &  STFT & L1 &  Mel &  STFT \\
    \midrule
    Alto        & 4.3 & 0.5 & -22.0 & -55.8 & -40.1 & -10.4 \\
    Bass        & 14.9 & 10.8 & -20.0 & -58.7 & -41.3 & -9.9 \\
    Lead Vocal  & 4.0 & 0.5 & -33.1 & -53.6 & -39.1 & -9.9 \\
    Soprano     & 1.4 & -8.0 & -41.0 & -52.3 & -39.6 & -12.0 \\
    Tenor       & 4.2 & -0.3 & -34.2 & -59.1 & -41.3 & -9.9 \\
    Vocal Perc. & 7.2 & -1.4 & -35.1 & -57.2 & -41.4 & -10.2 \\
    \bottomrule
  \end{tabular}
  \caption{Loss ablation study of \model{} under the subset condition on the JaCappella test split. We train solely with either L1, Mel, or STFT loss. We report per-stem SI-SDRi for active references and RMS (dBFS) for silent stems.}
  \label{tab:subset_snri_rms_ablation}
\end{table}

\begin{figure*}[h!]
    \centering
    \includegraphics[width=0.85\linewidth]{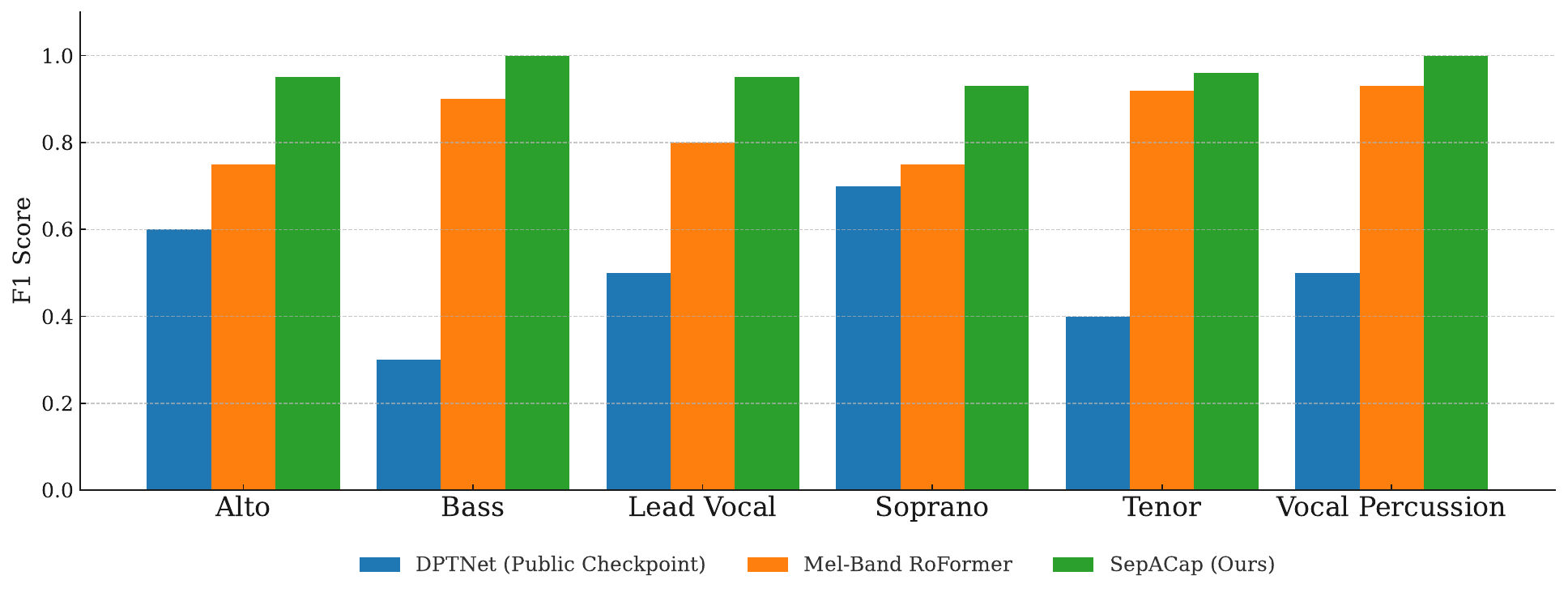}
    \caption{We test the ability of models to predict whether a stem is present in a mix. We report F1 scores for stem detection on the test split of the augmented JaCappella dataset. The results show per-stem detection performance for DPTNet, Mel-Band RoFormer, and SepACap, where higher is better in the interval. Our proposed model \model{} achieves the best overall performance in stem detection.}
    \label{fig:placeholder}
\end{figure*}

\textbf{Setup.}
We use the JaCappella dataset~\cite{nakamura2023jacappella} to train and evaluate \model{} and baseline approaches. The dataset provides Japanese a cappella music and their 6 corresponding stems (Alto, Bass, Lead Vocal, Soprano, Tenor, and Vocal Percussion). Using the data augmentation strategy discussed in \cref{sec:method}, we increase the dataset size to 105k samples up from the original 35 music clips. In terms of duration, the augmentation strategy increases the dataset duration from 0.57 hours to 145 hours.

We train \model{} on 4-second snippets of the augmented JaCappella dataset. The model is designed to separate up to six stems directly in the waveform domain. For training, we adopt the composite loss setup described in \cref{sec:method}, which combines time-domain and spectral reconstruction objectives. The spectral loss operates at three STFT window lengths (512, 1024, and 2048), and the mel loss spans seven bins (5, 10, 20, 40, 80, 160, and 320) with window lengths of 32, 64, 128, 256, 512, 1024, and 2048, respectively. 

Additionally, we conduct three ablation studies where we only use one loss per model training. From this we found the importance of each loss for the composite objective. As shown in~\cref{tab:sisdr_ablations} and~\cref{tab:subset_snri_rms_ablation}, we found that L1 loss had the most impact on a successful separation and the STFT loss the least. For the final composite loss we use a weight of 0.3 for the STFT loss, as well as a weight of 0.7 for the Mel loss and a weight of 1.0 for the L1 loss.

Furthermore, as a baseline comparison, we train a Mel-Band RoFormer  model~\cite{wang2023mel} on the same dataset.\footnote{We train with \url{https://github.com/KimberleyJensen/Mel-Band-Roformer-Vocal-Model}} 
We evaluate SepACap model in two settings. First, we assess model performance when all stems are present. In this setting we can directly compare against prior work DPTNet~\cite{chen2020dual}, X-UMX~\cite{sawata2021all}, and MRDLA~\cite{nakamura2021time}, by reporting the SI-SDR-improvement (SDRi), defined as
\begin{equation}
    \text{SI-SDR-improvement}= \text{SI-SDR}_{\text{Pred}}- \text{SI-SDR}_{\text{Mixture}}
\end{equation}
and average the results across stems. Second, we evaluate the subset condition, where only a subset of stems is present in the mixture. In this setting, the model must both separate active sources and correctly output silence for absent sources. To capture this dual objective, we use two complementary metrics. When a reference signal is present, we report SDRi, which quantifies separation quality relative to the input mixture. When the reference stem is silent, SI-SDR is not meaningful; instead, we evaluate the model's ability to suppress spurious output by measuring the root-mean-square energy relative to full scale (RMS-DBFS), defined as
\begin{equation}
    \text{RMS-DBFS} = 20 \cdot \log_{10}  \sqrt{ \frac{1}{T} \sum_{t} x_t^2 + \varepsilon } ,
\end{equation}
where $x$ is the signal, $T$ is the length of the signal, and $\varepsilon$ is a small constant. This silence metric directly reflects the residual energy of the predicted signal and therefore serves as an indicator of how well the model avoids false positives in stems that should be silent.

\smallskip
\noindent \textbf{Evaluation.}
\begin{table}[t!]
  \centering
  \setlength{\tabcolsep}{3pt}
  \renewcommand{\arraystretch}{0.95}

  \begin{tabular}{lccccc}
    \toprule
    Stem & X-UMX&~%
    DPTNet&%
    MRDLA&%
    MBR& Ours\\
    \midrule
    Alto        & 13.5 & 11.9 & \textbf{14.7} & 6.3 & \underline{14.6} \\
    Bass        & 9.1  & \underline{19.7} & 10.2 & 17.8 & \textbf{23.2} \\
    Lead Vocal  & 7.5  & \underline{8.9}  & 8.7  & 0.7 & \textbf{13.0} \\
    Soprano     & 10.7 & 8.5  & \underline{11.8} & 4.5 & \textbf{13.1} \\
    Tenor       & 10.2 & \underline{14.9} & 11.3 & 10.3 & \textbf{17.0} \\
    Vocal Perc. & 21.0 & 21.9 & \underline{22.1} & 19.3 & \textbf{22.5} \\
    \bottomrule
  \end{tabular}
  \caption{Performance of models when all stems are present, measured on the test split of the JaCappella dataset. The metric is per-stem SDRi (higher is better) for each model. Best performance in bold, and second best performance is underlined. We find that \model{} performs the best overall.}
  \label{tab:sisdr_baselines_final}
\end{table}
In \cref{tab:sisdr_baselines_final} we observe the performance of the different methods. We find that \model{} outperforms previous approaches in 5 out of 6 stems (Bass, Lead Vocal, Soprano, Tenor, and Vocal Percussion) in SDRi even though only a fraction of the samples seen during training contain all stems simultaneously. Furthermore, the Mel-Band RoFormer  seems to significantly underperform at this task, which suggests that the time-frequency domain masking struggles to separate multiple sources contained in similar frequency bands.
 \begin{table}[h!]
  \centering
  \setlength{\tabcolsep}{4pt}
  \renewcommand{\arraystretch}{0.95}

  \begin{tabular}{lccc|ccc}
    \toprule
    & \multicolumn{3}{c}{SI-SDRi (dB)$\uparrow$} & \multicolumn{3}{c}{RMS (dBFS)$\downarrow$} \\
    \cmidrule(lr){2-4}\cmidrule(lr){5-7}
    Stem & DPTNet & MBR & Ours & DPTNet & MBR & Ours \\
    \midrule
    Alto        & -17.2 & 3.9 & \textbf{11.6} & -19.6 & -59.1 & \textbf{-92.7} \\
    Bass        & -30.8 & 15.5 & \textbf{20.4} & -33.7 & -70.8 & \textbf{-95.1} \\
    Lead V.  & -44.0 & 1.6 & \textbf{ 9.1} & -41.5 & -63.6 & \textbf{-91.9} \\
    Soprano     & -46.9 & 1.6 & \textbf{11.1} & -44.7 & -55.5 & \textbf{-85.6} \\
    Tenor       & -25.9 & 7.6 & \textbf{13.0} & -27.2 & -75.3 & \textbf{-95.7} \\
    Vocal P. & -32.4 & 18.3 & \textbf{18.4} & -33.6 & -73.1 & \textbf{-95.3} \\
    \bottomrule
  \end{tabular}
  \caption{Subset condition performance of DPTNet (publicly-available checkpoint), our trained Mel-Band RoFormer (MBR), and our proposed model \model{}, on the test split of the augmented JaCappella dataset. The per-stem SDRi is only reported when a reference signal is present, and RMS silence scores evaluate suppression quality for silent stems. Unsurprisingly, we find that DPTNet underperforms on the subset-stem task as it was only trained on full mixes. SepACap also significantly outperforms the Mel-Band RoFormer as it does not rely on frequency-based masking.}
  \label{tab:subset_snri_rms_baselines}
\end{table}
 
 The reported values for X-UMX, DPTNet, and MRDLA are taken from JaCappella~\cite{nakamura2023jacappella}. For the subset objective, the results in \cref{tab:subset_snri_rms_baselines} highlight a clear trade-off between the two models. \model{} generalizes especially well to this setting, as it produces fewer instances of bleed-through when stems are absent and can more effectively suppress inactive sources. However, this comes at the cost of introducing more audible artifacts in the reconstructed signals. In contrast, the Mel-Band RoFormer  yields cleaner outputs with fewer artifacts, but it frequently fails to fully suppress silent stems, leading to noticeable bleed-through between sources. This difference is consistent with the underlying model designs: the Mel-Band RoFormer  operates by masking unwanted frequency components, which prevents artifact creation but makes complete suppression of inactive signals difficult, whereas \model{} generates waveforms directly and is therefore more prone to artifact introduction.

Despite \model{}'s stronger overall performance in the subset setting, it does not always maintain consistent performance across all stems. In particular, both \model{} and Mel-Band RoFormer  struggle on the Alto stem compared to the all-stems setting. The lower quantitative results observed in this evaluation can often be attributed to failures in detecting a present stem and instead predicting silence, as illustrated in \cref{tab:subset_snri_rms_baselines}. Because the SI-SDR metric assigns large negative values in such cases, these errors disproportionately reduce the average scores.

\medskip
\noindent \textbf{Conclusion.} 
We introduced \model{}, a source separation model trained for a cappella mixtures. Evaluated on JaCappella, \model{} achieves state-of-the-art performance when all stems are present and substantially improves subset separation by suppressing inactive stems and reducing bleed-through compared to baseline approaches.

\clearpage
\vfill\pagebreak

\ninept
\bibliographystyle{IEEEbib}
\bibliography{strings,refs}

\end{document}